\documentclass[conference]{IEEEtran}
\IEEEoverridecommandlockouts
\usepackage{cite}
\usepackage[breaklinks=true]{hyperref}
\usepackage{breakcites}
\usepackage{amsmath,amssymb,amsfonts}
\usepackage{algorithmic}
\usepackage{graphicx}
\usepackage{textcomp}
\usepackage{xcolor}
\usepackage{subfig}
\usepackage{algorithm}
\usepackage{algorithmic}
\usepackage{amsmath}
\newcommand{\myindent}[1]{
\newline\makebox[#1cm]{}
}
\usepackage{listings}
\def\BibTeX{{\rm B\kern-.05em{\sc i\kern-.025em b}\kern-.08em
    T\kern-.1667em\lower.7ex\hbox{E}\kern-.125emX}}

\begin{document}

\title{Misconfiguration prevention and error cause detection for distributed-cloud applications\\
\thanks{\includegraphics[width=0.4cm]{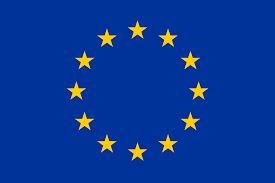} Funded by the European Union (TaRDIS, 101093006). Views and opinions expressed are however those of the author(s) only and do not necessarily reflect those of the European Union. Neither the European Union nor the granting authority can be held responsible for them.}
}

\makeatletter 
\newcommand{\linebreakand}{%
  \end{@IEEEauthorhalign}
  \hfill\mbox{}\par
  \mbox{}\hfill\begin{@IEEEauthorhalign}
}
\makeatother 

\author{\IEEEauthorblockN{Tamara Ranković}
\IEEEauthorblockA{\textit{Faculty of Technical Sciences} \\
University of Novi Sad \\
Novi Sad, Serbia \\
tamara.rankovic@uns.ac.rs}
\and
\IEEEauthorblockN{Filip Šiljić}
\IEEEauthorblockA{\textit{Faculty of Technical Sciences} \\
University of Novi Sad \\
Novi Sad, Serbia \\
siljic.sr18.2020@uns.ac.rs}
\and
\IEEEauthorblockN{Jovan Tomić}
\IEEEauthorblockA{\textit{Faculty of Technical Sciences} \\
University of Novi Sad \\
Novi Sad, Serbia \\
tomic.sw48.2019@uns.ac.rs}

\linebreakand

\IEEEauthorblockN{Goran Sladić}
\IEEEauthorblockA{\textit{Faculty of Technical Sciences} \\
University of Novi Sad \\
Novi Sad, Serbia \\
sladicg@uns.ac.rs}
\and
\IEEEauthorblockN{Miloš Simić}
\IEEEauthorblockA{\textit{Faculty of Technical Sciences} \\
University of Novi Sad \\
Novi Sad, Serbia \\
milos.simic@uns.ac.rs}
}

\maketitle

\begin{abstract}
Major software failures are reported to be due to misconfiguration.
As manual configuration is too error-prone to be deemed a reliable strategy for dynamic and complex systems, automated configuration management has become a standard.
Countermeasures against misconfiguration can be focused on prevention or, if failure already occurred, detection.
Configuration is often used as a broad term for any set of parameters or system states that dictate how an application will behave, but in this paper, we only focus on parameters consumed on process startup, usually from configuration files.
Our objective is to enhance configuration management processes in environments based on the distributed cloud model, a novel cloud model that allows dynamic allocation of strategically located resources.
The two mechanisms we propose are configuration validation using schemas and configuration version control with support for detecting differences between configuration versions.
Our solution reduces the risk of incorrect configuration as schemas prevent any non-compliant configuration from reaching applications.
However, if failure still occurs because the schema was incomplete or a valid configuration revealed existing software bugs, the version control system can precisely locate configuration changes that triggered the failure.
\end{abstract}


\section{Introduction}

Software programs are often parametrized so that they can exhibit different behaviors at runtime based on arguments passed to them.
We refer to those arguments as the program's configuration.
A common practice is to provide configuration parameter values inside configuration files that processes consume on start-up, or even later during execution if dynamic reconfiguration is supported.
We note that the term \emph{configuration} is frequently used in a much broader sense, covering also package installations, file permission settings, etc. \cite{baset2017usable}. However, through this paper, we will adhere to the definition we provided.

As software is getting more complex, so is its configuration process. 
Studies suggest that configuration interfaces are rapidly increasing over time \cite{zhang2021evolutionary}. 
This, combined with the fact that applications are becoming more sophisticated, consisting of numerous components and running in heterogenous environments, makes manual configuration inefficient and highly error-prone \cite{kostromin2020survey, karvinen2023configuration}.
Software configuration management (SCM) systems play a crucial role as they automate the configuration process and provide better control and visibility, as well as incident management and troubleshooting capabilities \cite{chinthapatla2024mastering}.

Even though SCM systems are becoming more advanced, misconfiguration remains a major cause of outages and software failures in production \cite{zhang2021evolutionary, tang2015holistic}. 
Zhang et al. \cite{zhang2021evolutionary} compiled a list of large-scale examples backing this argument.
They state that \emph{``misconfigurations were the second largest cause of service-level disruptions in one of Google's main production services. Misconfigurations also contribute to 16\% of production incidents at Facebook, including the worst-ever outage of Facebook and Instagram that occurred in March 2019''.}

Errors introduced by configuration modifications can be classified in the following way, according to \cite{tang2015holistic}: i) \emph{common errors}, such as typos or out-of-bound values, ii) \emph{subtle erros} leading to e.g. latency issues and butterfly effects, and iii) \emph{valid changes exposing software bugs}.
The solution we propose in this paper mainly focuses on the first class of errors, but can also aid in cause localization when it comes to the other two classes.

In this paper, we present mechanisms for misconfiguration prevention and error-cause detection for both infrastructure and applications running in the distributed cloud (DC). 
The two mechanisms we discuss are schema-based configuration validation and version control equipped with algorithms for calculating differences between configuration versions.

In Section \ref{s:rw}, we analyze existing solutions to dealing with configuration errors.
As DC is a novel cloud model, we introduce it in Section \ref{s:b}. 
Sections \ref{s:mp} and \ref{s:cecd} describe how we integrated configuration schemas and version control into DCs, respectively.
In Section \ref{s:i}, we present the system implementation.
Contributions and limitations of our work are discussed in Section \ref{s:d}, while Section \ref{s:c} provides directions for future work and concludes the paper.

\section{Related work}
\label{s:rw}

In this section, we discuss existing misconfiguration-related solutions.
In Section \ref{s:d} we will compare our work to the one presented here.
Questions we are most concerned with are:
\begin{itemize}
    \item How to prevent misconfiguration?
    \item When errors following configuration changes are detected, how to determine the root cause?
\end{itemize}

Facebook engineers take the \emph{configuration-as-code} approach, relying on Python and Thrift \cite{thrift} to define the structure and rules for programmatic manipulation of configuration, e.g. parameter value validation \cite{tang2015holistic}.
As their software undergoes thousands of configuration changes daily, they follow a thorough framework defined in \cite{tang2015holistic}:
\begin{itemize}
    \item \emph{Compiler validation} - As configuration source code gets compiled down to JSON files, any compile-time error can be detected at an early stage, preventing it from ever reaching applications.
    \item \emph{Code reviews} - Both source code and generated JSON files reside in git, and are subjected to code reviews, just like the regular code is.
    \item \emph{Manual testing} - New configuration gets submitted to several test or production servers so that the application's behavior can be observed.
    \item \emph{Canary testing} - Following the submitted specification, the new configuration is applied to a percentage of live production servers. The specification also contains criteria to determine if the tests passed or failed, which helps trigger the rollback.
\end{itemize}

Elektra \cite{raab2020unified} is a configuration storage and file-generating system that aims to solve the problem of dealing with different configuration file formats, such as JSON, YAML, HOST, INI, etc.
It provides a simple unified API that lets users mount their files and manage parameters, not having to worry about files' syntactic validity.
The authors report their solution leads to higher productivity when performing configuration management tasks.

ConfigValidatior \cite{baset2017usable} is a rule engine that consumes configuration files in different formats, normalizes that data, and produces output based on user-defined rules.
The rules are written in a custom language defined for this tool's purposes.
It is worth noting that ConfigValidator can be treated as misconfiguration prevention, rather than detection, software only if the input files are stored in some staging location and the validation is performed before their submission to applications.

In the same paper, it is discussed how configuration validation tools can be divided into rule- and inference-based ones.
As inference systems relying on machine learning have an inherent error delta, they are rarely seen in production environments \cite{baset2017usable}.

The idea behind ctests \cite{sun2020testing} is to parametrize tests with configuration parameters and then run them passing production configuration as an input.
This approach could help with eliminating errors before they reach production. 
However, ctests' effectiveness is highly dependent on the quality of the tests. 
Also, implementing all tests from scratch would require substantial up-front effort, which is less of a problem for mature systems that already contain them \cite{sun2020testing}.
It is emphasized by the authors that ctests can't necessarily determine the error's root cause, but can localize it.

\section{Distributed clouds}
\label{s:b}

In this section we will provide a brief overview of the DC model, after which we will discuss how configuration management was integrated into an open-source implementation of the DCs, \emph{Constellations (c12s)}\footnote{c12s source code: \url{https://github.com/c12s}}.
Research presented in this paper builds on top of the mentioned system, providing it with misconfiguration prevention and cause detection mechanisms.

\subsection{Overview}

The cloud as we know it allows flexible allocation of resources based on the current requirements, which makes it a widely adopted solution.
However, use cases from recent years have shown us the shortcomings of the cloud’s centralized resource organization \cite{ferrer2019towards}. Examples include systems that need to process large amounts of data in real-time or comply with legal constraints on where user data can be stored or processed.
To solve these problems, new cloud paradigms have emerged that try to extend or completely replace the traditional cloud \cite{bonomi2012fog, shi2016edge}.
Among them is the DC model \cite{simic2021towards, simic2021infrastructure, simic2024hierarchical}. It places a DC layer between the cloud and its clients. 
The primary idea behind DCs is that users should be able to dynamically allocate resources that are placed strategically, for example in proximity to their applications’ clients.

When it comes to DC architecture, we can observe it from two different perspectives, physical and logical.
First, there is a control plane in the cloud managing physical infrastructure in the DC layer.
Users provision infrastructure and run their applications on it by interacting with the control plane.
All users are members of an organization. Therefore, when they provision infrastructure, it will be deemed as temporarily owned by that organization, and accessible by other members.
Looking from a logical perspective, we can say that every organization has a resource pool available. That pool can be partitioned using namespaces, isolating physical resources, applications, and data.

\subsection{Configuration management in the distributed cloud}

Applications running in the DC can be supplied with configuration parameter values via the configuration management subsystem. Its primary responsibilities include configuration storage, manipulation, and dissemination.
First, we will see the configuration model it supports, and then a simple dissemination workflow.

When specifying and consuming configuration, we distinguish between \emph{1) standalone configuration}, and \emph{2) configuration groups}.
Each standalone configuration represents a named key-value pair set, while a configuration group is a named collection of arbitrarily many named key-value pair sets.
Configurations are organization-scoped resources, i.e. every configuration defined is owned by an organization.
For audit purposes, both configuration types are versioned, meaning every configuration instance has a \emph{version} attribute, and is immutable.
It follows that we can uniquely identify every configuration by a combination of its owner organization, name, and version.
Both keys and values of configuration pairs are treated as strings.
Users can program their applications to interpret the values as any data type but are responsible for serialization when declaring the configuration and deserialization when the application reads the values.

User-defined configuration is not made instantly available to DC applications, but rather stored only in the control plane and needs to be explicitly disseminated to a subset of nodes from the organization's node pool.
Every dissemination request should declare a unique identifier of the configuration, namespace, and label-based query. The configuration, if found, will be put in the specified namespace and by default visible only to applications inside that namespace.
This behavior can be overridden with access control policies supported by the platform, which were introduced by Ranković et al. in \cite{rankovic2024access}.
The query's purpose is to select nodes from the pool that meet certain conditions.
A node will match the query if its label values satisfy the constraints set by the query.
But what is a label? It is a key-value pair describing node's relevant attributes, such as software and hardware specification, and location.
Every node has an initial label set that can be modified later on.

\section{Misconfiguration prevention}
\label{s:mp}

Invalid parameter names or values can be a cause of misconfiguration that is relatively easily preventable.
Aside from manual reviews, it is possible to define rules that the configuration would be automatically validated against before it gets disseminated or even stored in the system.
One way of expressing those rules is through a data schema.

A JSON schema \cite{json} is a standardized way of describing the expected format of JSON documents.
It can articulate various rules governing data validation and organization, including specifying required fields, defining permissible data types, setting minimum and maximum values, validating patterns with regular expressions, and enforcing hierarchical relationships between elements.
Additionally, JSON schemas can express conditional logic and dependencies.

A YAML schema \cite{yaml} functions similarly to a JSON schema but is tailored specifically for the YAML language.
It can express any rule a JSON schema can, but additionally has features such as tags, property ordering, serialization styles, and examples.

As a higher level of YAML schema expressivity doesn't benefit our use case, we chose to rely on the JSON schema for validation.
However, as YAML is more compact and perhaps readable, we let users specify the schema in that format and then translate it to JSON internally.

We allow schema definition for both standalone configuration and configuration groups.
An example of configuration instances of the two types serialized to YAML is displayed in Listing \ref{l:yaml}.
As the system will translate configurations to the JSON equivalent of this structure before validating it, users should define their schemas accordingly.
Even though we eventually serialize everything to JSON, users are not aware of it and must know only the YAML representation.

\begin{lstlisting}[
    caption=Configurations serialized to YAML,
    label=l:yaml,
    float=!t,
    floatplacement=!t,
  ]
    # standalone configuration
    param1: value1
    param2: value2
    param3: value3

    # configuration group
    config1:
        param1: value1
        param2: value2
    config2:
        param3: value3
        param4: value4
\end{lstlisting}

Outlined below is the complete list of schema-related operations:
\begin{itemize}
    \item\textbf{Schema Storing} - Despite being accepted in YAML format, the provided string, translated to JSON, must be a valid JSON schema for the reasons we previously discussed. Additionally, a name, version, and organization for the schema must be specified.
    \item\textbf{Schema Retrieval} - A schema can be retrieved by specifying its unique identifier, which is the combination of the name, version, and organization.
    \item\textbf{Schema Version History} - This operation can be used to retrieve all schemas under the given organization and schema name.
    \item\textbf{Schema Deletion} - As for the retrieval, a unique identifier must be stated.
    \item\textbf{Configuration Validation} - New configurations can be validated against previously saved schemas. The configuration will be accepted by the system only if the validation terminates successfully.
\end{itemize}

\section{Configuration error cause detection}
\label{s:cecd}

Application failure due to configuration changes can be caused by invalid configuration or by previously undetected bugs in the application.
In both cases, what triggered the errors is a difference between the last and current configuration.
Thus, the first step of the troubleshooting process is to determine the delta from the latest stable configuration and use it as a starting point for detecting the root cause of the errors.
Our version control system (VCS) aids in this particular task by locating the diff between the two configurations.

As stated in Section \ref{s:b}, we version both standalone configuration and configuration groups by labeling every configuration instance with a \emph{version} attribute.
Users may choose any string value meaningful to them to be a version.
Configuration instances are immutable, meaning the parameter values can't be changed for an existing configuration version.
If users wish to change them, they must declare a new version.

This design decision was made so that no configuration can get hidden by updates.
If it weren't for it, the following scenario could be possible:
\begin{itemize}
    \item A configuration version gets declared.
    \item It gets disseminated and consumed by an application.
    \item The configuration with that version gets updated.
    \item The application that had previously consumed the configuration starts misbehaving.
    \item During auditing, the configuration is requested from the system.
    \item The system returns the updated configuration, and not the one the application is using.
\end{itemize}
Although it is possible to store every change of the same version, it then requires comparing update and dissemination times to determine which configuration instance the application consumed.
We deem this approach unnecessarily complex and error-prone, especially when updates are frequent.

\begin{algorithm}[t]
    \caption{ParamSetDiff}
    \label{a:1}
    \begin{algorithmic}[1]
        \renewcommand{\algorithmicrequire}{\textbf{Input:}}
        \renewcommand{\algorithmicensure}{\textbf{Output:}}
        \REQUIRE ref, target NamedParamSet
        \ENSURE  diffs []Diff
        \FOR {$param$ in $target$}
            \IF {($param$ not in $ref$)}
                \STATE diff = Addition\{key: param.key,
                    \myindent{0.3} value: param.value\}
                \STATE diffs.append(diff)
            \ELSIF {($param.value \ne ref[param.key].value$)}
                \STATE diff = Modification\{key: param.key, 
                    \myindent{0.3} oldValue: ref[param.key].value,
                    \myindent{0.3}  newValue: param.value\}
                \STATE diffs.append(diff)
            \ENDIF
        \ENDFOR
        \FOR {$param$ in $ref$}
            \IF {($param$ not in $target$)}
                \STATE diff = Deletion\{key: param.key,
                    \myindent{0.3} value: param.value\}
                \STATE diffs.append(diff)
            \ENDIF
        \ENDFOR
        \RETURN diffs
    \end{algorithmic}
\end{algorithm}

\begin{algorithm}[h]
    \caption{ConfigGroupDiff}
    \label{a:2}
    \begin{algorithmic}[1]
        \renewcommand{\algorithmicrequire}{\textbf{Input:}}
        \renewcommand{\algorithmicensure}{\textbf{Output:}}
        \REQUIRE ref, target ConfigGroup
        \ENSURE  diffs map[string, []Diff]
        \FOR {$nps$ in $target.namedParamSets$}
            \IF {($nps$ not in $ref.namedParamSets$)}
                \FOR {$param$ in $nps$}
                    \STATE diff = Addition\{
                        \myindent{0.3} key: param.key, value: param.value\}
                    \STATE diffs[nps.name].append(diff)
                \ENDFOR
            \ELSE
                \STATE npsDiffs = ParamSetDiff(
                    \myindent{0.3}  ref.namedParamSets[nps.name], nps)
                \STATE diffs[nps.name].append(npsDiffs)
            \ENDIF
        \ENDFOR
        \FOR {$nps$ in $ref.namedParamSets$}
            \IF {($nps$ not in $target.namedParamSets$)}
                \FOR {$param$ in $nps$}
                    \STATE diff = Deletion\{key: param.key,
                        \myindent{0.3} value: param.value\}
                    \STATE diffs[nps.name].append(diff)
                \ENDFOR
            \ENDIF
        \ENDFOR
        \RETURN diffs
    \end{algorithmic}
\end{algorithm}

The first operation our VCS offers for detecting configuration changes is \textbf{timelining}.
The client specifies the organization and name of the standalone configuration or configuration group, and the system composes a chronologically ordered history of configuration versions.
When parameter sets are reasonably short, this mechanism enables straightforward identification of modifications that occur through time.
However, in complex configurations, especially groups that may comprise any number of lengthy parameter sets, timelining can easily become insufficient.
For those scenarios, we developed a \textbf{diff} operation.

\begin{lstlisting}[
    caption=Example of the configuration group diff,
    label=l:e,
    float=!t,
    floatplacement=!t,
  ]
# Reference configuration group
{
    "name": "group1",
    "version": "1.0.0",
    "organization": "orgA",
    "namedParamSets": [
        {
            "name": "configA",
            "params": {
                "pA": "vA"
            }
        },
        {
            "name": "configB",
            "params": {
                "pB": "vB"
            }
        }]
}
# Target configuration group
{
    "name": "group1",
    "version": "1.0.1",
    "organization": "orgA",
    "namedParamSets": [
        {
            "name": "configA",
            "params": {
                "pA": "vA2"
            }
        },
        {
            "name": "configC",
            "params": {
                "pC": "vC"
            }
        }]
}
# diff
{
    "configA": [
        {
            "type": "Modification",
            "key": "pA",
            "oldValue": "vA",
            "newValue": "vA2"
        }],
    "configB": [
        {
            "type": "Deletion",
            "key": "pB",
            "value": "vB"
        }],
    "configC": [
        {
            "type": "Addition",
            "key": "pC",
            "value": "vC"
        }]
}
\end{lstlisting}

As other VCSs do, our diff operation too detects parts that differ between two snapshots of the data.
In our case, the operation is specialized for data to have the format of the configuration.
The system offers two options, finding diffs between two arbitrary standalone configurations or two configuration groups.
For both, the organization, name, and version of the reference configuration must be specified, as well as the organization, name, and version of the target configuration.
The operation will calculate all changes introduced in the target configuration when compared to the reference configuration.
There is no constraint that both configurations must have the same name and organization, but we think the diff operation is most usable when that is the case.

A diff between two standalone configurations is a list of atomic diffs. 
An atomic diff can be one of the following: \emph{i) addition}, \emph{ii) delition} or \emph{iii) modification}.
The addition diff denotes that a new parameter was introduced in the target configuration, deletion that an existing one was removed, and modification that the value of an existing parameter was changed.
A diff between two configuration groups is a map of diff lists. Keys in the map are a union of configuration names from both groups.
The algorithm \ref{a:1} calculates the diff between named parameter sets.
When comparing standalone configurations, we directly invoke the implementation of this algorithm.
In the case of configuration groups, we use the algorithm \ref{a:2}.
In Listing \ref{l:e}, we can see an example of the diff between configuration groups.

\begin{figure}[t]
    \center
    \includegraphics[width=0.8\columnwidth]{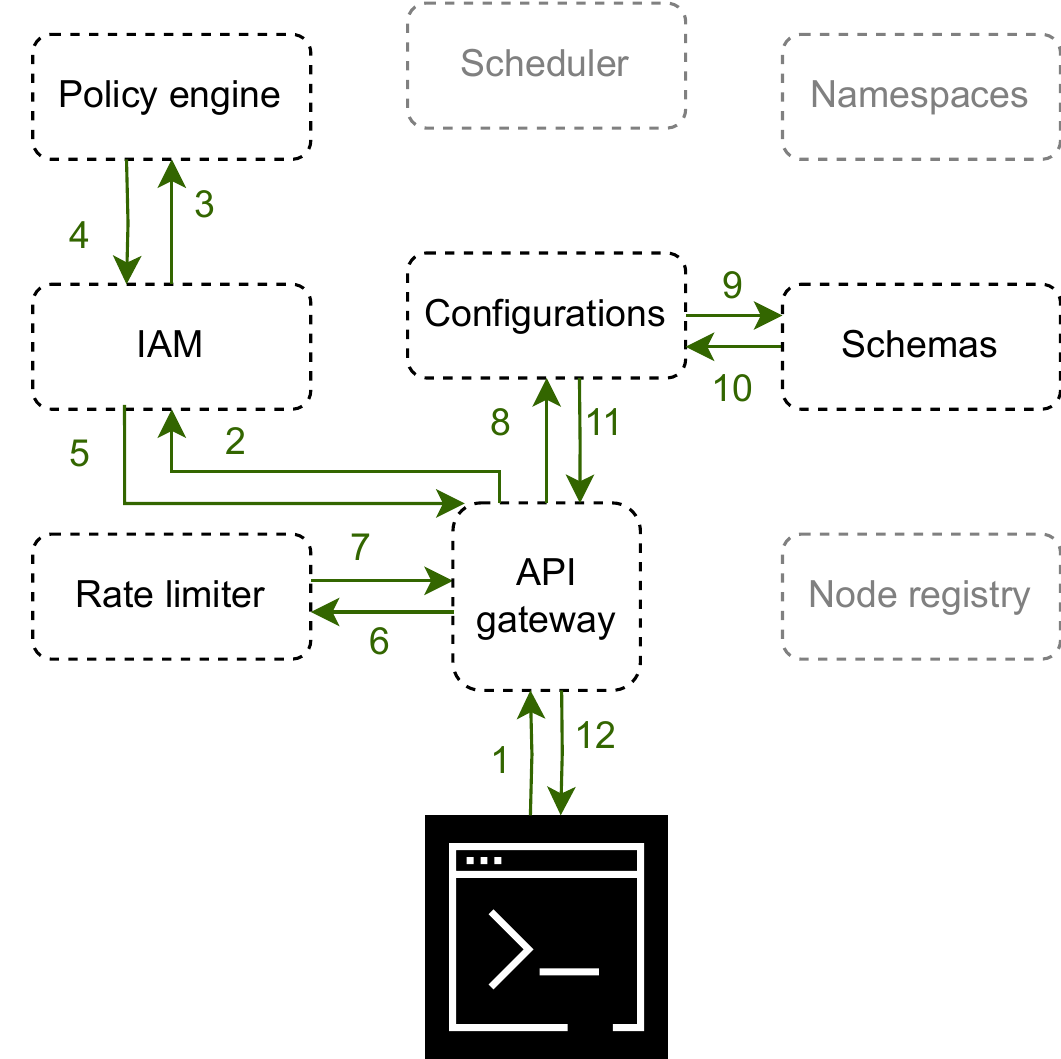}
    \caption{c12s control plane}
    \label{fig:a}
\end{figure}

\section{Implementation}
\label{s:i}

The control plane of the \emph{c12s} platform is implemented as a microservice application.
We extended it with a service dedicated to schema management, while the VCS was added to an already-existing service for configuration management.
Both are implemented in Go programming language and expose a gRPC\footnote{gRPC: \url{https://grpc.io/}} API to other services.
The etcd\footnote{etcd: \url{https://etcd.io/}} key-value store was used for persisting the data.

Figure \ref{fig:a} presents the architecture of the control plane, as well as the flow of a successful request that adds a new configuration to the system.
The configuration is submitted to a \emph{CLI tool}\footnote{CLI tool source code: \url{https://github.com/c12s/cockpit}} that interacts with an \emph{API gateway}\footnote{API gateway source code: \url{https://github.com/c12s/lunar-gateway}}.
The gateway performs, among other things, authentication and rate limiting.
In doing so, it directly communicates with \emph{Identity and Access Management (IAM)}\footnote{IAM service source code: \url{https://github.com/c12s/apollo}} and \emph{Rate limiter}\footnote{Rate limiter service source code: \url{https://github.com/c12s/heliosphere}} services.
IAM service receives the user's authentication token and, if it is valid, requests an authorization token from the \emph{Policy engine}\footnote{Policy engine source code: \url{https://github.com/c12s/oort}}.
Upon getting the authorization token, the gateway propagates it together with the request to the \emph{Configurations}\footnote{Configurations service source code: \url{https://github.com/c12s/kuiper}} service.
If the user is authorized to perform the initiated action, the configuration will first be validated against a schema by the \emph{Schemas}\footnote{Schemas service source code: \url{https://github.com/c12s/quasar}} service.
If no errors are detected, the configuration is stored and a confirmation is sent back to the user.

\section{Discussion}
\label{s:d}

Our solution allows users of the \emph{c12s} DC platform to reduce the risk of misconfiguration by leveraging schemas and a custom-designed VCS.
Configuration is declared and fetched by applications through a simple API, which removes concerns about the syntactic correctness of configuration files.
For validation rule language, we opted for the widely adopted YAML schema, and not a DSL the users would have to learn before using the platform.
Configuration is validated upon definition so that it's not possible to disseminate it if any error is detected.
Still, failures caused by configuration errors unforeseen by the schema or bugs triggered by valid configuration remain.
In these cases, the VCS can determine the exact changes that were made from the last stable configuration.
Those are likely the initiators of the failure.
What differentiates our VCS from other general-purpose VCSs like git, often used for versioning configuration, is the fact that it is specialized for the configuration format the platform uses.
Consequently, it can pinpoint the modifications by looking at the parameters, and not just lines of arbitrary text in a file.

A significant limitation of applications programmed to run in DCs is that they are currently not portable to other environments as they rely on making API calls to get their configuration.
We can overcome this problem by providing an option to serve configuration in the form of files stored in a predefined location.
Also, even though it is possible to perform canary tests by disseminating configuration only to a subset of nodes, this process can be streamlined by introducing additional out-of-the-box testing primitives.

Before incorporating our solution into a real-life setting, its effectiveness should be analyzed for particular use cases.
We propose the following testing strategies, aimed to guide the users in estimating the advantages our tools could provide.
The number of incorrect configurations provisioned with and without relying on a schema can be a good indicator of its usefulness.
Furthermore, the time required to construct a correct schema is also an important factor as schema management represents additional effort invested.
Regarding the VCS, our diff algorithms were designed to have both precision and recall of value 1 for detecting modifications in configurations and configuration groups.
In comparison, git's default algorithm, which would also score the maximum recall, would result in variable precision determined by the input structure.
For example, reordering parameters is not deemed configuration modification, but git would treat it as one.
So, the precision itself could be measured for different configurations. However, its direct effect on VCS usability may be unclear.
We hypothesize that lower precision leads to more time needed by the user to infer what changes were made.
But, until we conclusively determine the correlation, it could be more valuable to measure that time for outputs generated by our VCS and, for example, git.

\section{Conclusion}
\label{s:c}

This paper presents mechanisms for misconfiguration prevention and error-cause detection meant to be used in a DC environment.
Configurations get validated against a schema before they are ever stored in the system, so there is no possibility of applying configuration non-compliant with the schema.
Additionally, failures triggered by configuration that reached applications can be examined with the help of a custom-built VCS that detects changes between configuration versions.

Our vision is to extend the set of measures that can be taken to eliminate misconfiguration, including support for straightforward testing.
Furthermore, our solution is going to be extended with alerting and rollback mechanisms reacting to errors following configuration changes.
Additionally, as we plan to integrate the execution of user-specified shell command sequences into the configuration process, we must develop countermeasures against misconfiguration in those situations.

\bibliographystyle{bib/IEEEtran}
\bibliography{bib/refs}

\end{document}